\documentstyle[prl,aps,twocolumn,epsf]{revtex}
\def\break#1{\pagebreak \vspace*{#1}}
\begin{document}

\draft

\title{Supersymmetric strictly isospectral FRW models for zero factor ordering}

\author{H.C. Rosu 
and J. Socorro
}

\address{ 
{Instituto de F\'{\i}sica de la Universidad de Guanajuato, Apdo Postal
E-143, Le\'on, Gto, M\'exico} 
 }

\maketitle
\widetext

\begin{abstract}

We apply the strictly
isospectral technique of standard supersymmetric quantum
mechanics to the $Q=0$ factor ordered Wheeler-DeWitt equation for
the closed Friedmann-Robertson-Walker (FRW) minisuperspace model. The
resulting strictly isospectral one-parameter families of both FRW
cosmological potentials and ``wavefunctions of the universe" are exhibited
with relevant plots.

\end{abstract}

\pacs{PACS numbers: 98.80.Hw, 11.30Pb, 04.60.Kz \hspace{1cm}
 gr-qc/9606030
\hspace{2.5cm}
 Nuovo Cimento B 113, 683-689 (1998)}

\narrowtext


Bianchi cosmologies \cite{R}
are cosmological models which are currently
used to illustrate possible ways of finding exact solutions to the canonical
constraints in quantum cosmology. In the last decade
the number of exact solutions has been continuously increased by applying
various formalisms \cite{os}. The simplest of all cosmologies is
the FRW model because of its full one-dimensional character.
The goal of the present work is to apply
the strictly isospectral scheme based on the general Riccati solution
\cite{M84,N84}, which is quite well known in
nonrelativistic one-dimensional supersymmetric quantum mechanics,
to the $Q=0$ factor-ordered Wheeler-DeWitt
(WDW) equation corresponding to the closed FRW cosmological model.
Thus, we will be able to display a one-parameter class of strictly
isospectral cosmological FRW solutions and to present the
wavefunctions of the universe for that case.

FRW models belongs to the diagonal Bianchi type IX models in the
class A models \cite{R}.
Therefore we recall briefly the canonical formulation in the
Arnowitt-Deser-Misner (ADM) formalism
of the latter ones. The metric has the form
\begin{equation} \label{1}
\rm{ ds^2= -dt^2 + e^{2\alpha(t)}\, (e^{2\beta(t)})_{ij}\, \omega^i \,
\omega^j,}
\end{equation}
where $\alpha(t)$ is a scalar function
and $\rm \beta_{ij}(t)$ a 3x3 diagonal
matrix, 
$\rm \beta_{ij}= diag(x+ \sqrt{3} y,x- \sqrt{3} y, -2x)$,
$\rm \omega^i$ are one-forms which characterize each cosmological Bianchi
type model, obeying
$\rm d\omega^i= {1\over 2} C^i_{jk} \omega^j \wedge \omega^k,$
$\rm C^i_{jk}$ being the structure constants of the corresponding invariance
group.

 The ADM action has the form
%
\begin{equation} \label{2}
\rm {I=\int (P_x dx+ P_y dy + P_{\alpha} d\alpha -  N {\cal H}_\perp) dt,}
 \end{equation}
%
where the P's are canonically conjugated momenta, N is the lapse function
and
\begin{equation} \label{3}
{\cal H}_\perp= \rm e^{-3\alpha}\left(-P^2_\alpha+ P^2_x +P^2_y +
e^{4\alpha}V(x,y) \right)~.
\end{equation}
$\rm e^{4\alpha}\, V(x,y)= U(q^{\mu})$ is the potential term of the
cosmological model under consi\-de\-ra\-tion. 
The WDW equation for
\break{1.34in}
these models is achieved by
replacing $\rm {P_{q^{\mu}}}$
by $\rm{ -i \partial_{q^{\mu}}}$ in Eq.~(3),
with $\rm{ q^{\mu}=(\alpha, x,  y)}$.
 The factor $\rm{ e^{-3\alpha}}$ may be
factor-ordered
with $\rm {P_{q^{\mu}}}$ in many ways.
 Hartle and Hawking \cite{hh} have
suggested an almost
general factor ordering, which in this
case would order
$\rm{ e^{-3\alpha} P^2_{q^{\mu}}}$ as
%
\begin{equation} \label{4}
\rm{ - e^{-(3- Q)\alpha}\, \partial _{\alpha} e^{-Q\alpha} \partial _{\alpha}
=
- e^{-3\alpha}\, \partial^2_{\alpha} + Q\, e^{-3\alpha} \partial _{\alpha},}
\end{equation}
where $Q$ is any real constant. In the following, we shall work in the $Q=0$
factor ordering in which the WDW equation reads
%
\begin{equation} \label{5}
\rm {\Box \, \Psi - U(q^{\mu}) \, \Psi =0,}
\end{equation}
and by means of the Ansatz
$\rm \Psi(q^{\mu}) = A e^{\pm \Phi}$
one gets
\begin{equation}  \label{6}
  \pm A {\Box \, \Phi}  
+ A [ (\nabla \Phi)^2 - U] = 0,
\end{equation}
where  $\rm \Box = G^{\mu \nu}{\partial ^2\over \partial q^{\mu} \partial
 q^{\nu}}$,
$\rm (\nabla) ^2= -({\partial \over \partial \alpha})^2
+({\partial \over \partial x})^2 +
({\partial \over \partial y})^2$, and
$\rm G^{\mu \nu}= diag(-1,1,1)$. 

For the FRW model there is only the $\alpha$ coordinate and
the corresponding potential is $U=-e^{4\alpha}$.
Then the WDW equation can be written as a one-dimensional ordinary
differential equation
\begin{equation} \label{8}
 \left( {d \Phi}\over {d \alpha}  \right) ^2
   \pm   \left({d ^2 \Phi}\over {d \alpha ^2} \right)
=e^{4\alpha}
\end{equation}
and allows one to think of the Witten supersymmetric procedure
(for a recent review see \cite{csk})
since the change of function $\Phi=\pm\ln y$ produces a zero-energy
one-dimensional
Schr\"odinger equation, whereas the change $\frac{\partial \Phi}
{\partial \alpha}= \pm{\cal W}$
leads to a Riccati equation for ${\cal W}$, which can be identified
with the superpotential in the Witten procedure.
The one-dimensional Schr\"odinger equation reads
\begin{equation} \label{88}
-y^{''}+e^{4\alpha}y=0~,
\end{equation}
with the second derivative taken with respect to the $\alpha$ coordinate,
and possessing as general
solution a superposition of modified Bessel functions
$Z_0(i\frac{1}{2}e^{2\alpha})= C_1 I_0(\frac{1}{2}e^{2\alpha})+
C_2K_0(\frac{1}{2}e^{2\alpha})$.
Let us consider separately the two particular cases as follows:

(i) $Z_0=K_0$; This case corresponds asymptotically to
an Ansatz of the type $\Psi=Ae^{-\Phi _{1}}$
for the
``wavefunction of the universe", with
$\Phi _{1} =\frac{1}{2}e^{2\alpha}+ \alpha$.
Indeed, asymptotically ($\alpha \gg 1$)
\cite{bow}
\begin{equation} \label{13}
K_0(\frac{1}{2}e^{2\alpha})\approx e^{-\frac{1}{2}e^{2\alpha}-\alpha}~.
\end{equation}

(ii) $Z_0=I_0$; Again asymptotically,
it corresponds to an Ansatz of the type
$\Psi=Ae^{+\Phi _{2}}$ for the 
`wavefunction of the universe', with
$\Phi _{2} =\frac{1}{2}e^{2\alpha}-\alpha$ since for $\alpha \gg 1$
\cite{bow}
\begin{equation} \label{15}
I_0(\frac{1}{2}e^{2\alpha})\approx e^{\frac{1}{2}e^{2\alpha}-\alpha}~.
\end{equation}

We shall now apply the technique of strictly isospectral
potentials to the FRW model. This technique
is known since about a decade in one-dimensional
supersymmetric quantum mechanics and usually
requires nodeless, normalizable states of Schr\"odinger equation.
However, Pappademos, Sukhatme, and Pagnamenta \cite{psp}
showed that the strictly isospectral construction can be also
performed on nonnormalizable states.
Thus, using the nonnormalizable modified Bessel function $Z_0$  the
strictly isospectral construction says that the isospectral potentials for
the FRW cosmological model are given by
\begin{eqnarray} \nonumber
S_{iso}(\alpha;\lambda) &=&
S^{-}(\alpha)-2[\ln({\cal J} _0+\lambda)]''\\ \label{s2}
&=&
S^{-}(\alpha)-\frac{4Z_0(\alpha)Z_0^{'}(\alpha)}{{\cal J} _0+\lambda}+
\frac{2Z_0^4(\alpha)}{({\cal J} _0+\lambda)^2}~,
\end{eqnarray}
where $S^{-}(\alpha)=e^{4\alpha}$ is the Schr\"odinger-FRW potential,
$\lambda$ is a real parameter and
\begin{equation} \label{10}
{\cal J} _0(\alpha)\equiv\int_{-\infty}^{\alpha}Z_0^2(i\frac{1}{2}e^{2y})dy~.
\end{equation}
The common argument for the supersymmetric strictly isospectral method is
as follows. One can factorize the FRW one-dimensional
Schr\"odinger equation with the operators
 $A=\frac{d}{d\alpha}+{\cal W}(\alpha)$ and
$A^{\dagger}=-\frac{d}{d\alpha}+{\cal W}(\alpha)$,
 where the superpotential function
is given by ${\cal W}=-Z_0^{'}/Z_0$. Then, in the Witten scheme,
the potential and superpotential enter an initial `bosonic' Riccati equation
$S^{-}={\cal W} ^2-{\cal W} ^{'}$. At the same time, one can build
a `fermionic' Riccati equation $S^{+}={\cal W} ^2+{\cal W} ^{'}$,
 corresponding to a
`fermionic' Schr\"odinger equation for which the operators
 $A$ and $A^{+}$
are applied in reversed order.
 Thus, the `fermionic' Schr\"odinger- FRW potential
is found to be
 $S^{+}=e^{4\alpha}-2 (\frac{Z_0^{'}}{Z_0}
)^{'}$. This potential does not have
the $Z_0$ solution as a `wavefunction of the universe'.
 However, it is possible
to reintroduce the $Z_0$ solution into the spectrum, by
using the general superpotential solution of the `fermionic' Riccati
equation. The general Riccati solution reads
\begin{equation} \label{11}
{\cal W} _{gen}={\cal W}(\alpha) +
 \frac{d}{d\alpha}\ln [{\cal J} _0(\alpha)+\lambda]~.
\end{equation}
The way to obtain Eq.~(13) is well-known \cite{M84,N84} and will not
be repeated here.
From it one can easily get Eq.~(11). The one-parameter family of
`wavefunctions of the universe' differs from the initial one, being the
$\lambda$-depending quotient
\begin{equation} \label{12}
Z_{gen}(\alpha;\lambda)= \sqrt{\lambda (\lambda +1)}
\frac{Z_0(i\frac{1}{2}e^{2\alpha})}{{\cal J} _0+\lambda}~,
\end{equation}
though in the limit $\lambda \rightarrow \pm\infty$ one recovers the initial
solution. In Eq.~(14) we have included the
factor $\sqrt{\lambda(\lambda+1)}$,
however not for normalization reasons as in supersymmetric quantum mechanics
\cite{priv} (the wavefunctions are still nonnormalizable) but in order to
discuss the so-called Pursey limit \cite{P} $\lambda =0$, and
the Abraham-Moses limit \cite{AM} $\lambda =-1$ (see figure captions).
In order to get a better feeling of what is going on we have plotted
the strictly isospectral FRW
cosmological potentials and wavefunctions for $Z_0=I_0$ and $Z_0=I_0-K_0$
and two values of the $\lambda$- parameter, see Figures 1, 2 and 3,
confirming in this cosmological context some known facts of the
supersymmetric strictly isospectral
construction in nonrelativistic quantum mechanics \cite{csk}.
What we actually
did was to calculate the integral and all the other quantities
in the interval $(-c, \alpha)$,
where $c$ is a sufficiently big positive number, while $\alpha$ has been
chosen a not too big positive number. In other words, we have
a particle (the FRW universe) in a special type of box, that we call a
Bessel strictly isospectral box.
For $I_0$, the integral ${\cal I} _{0}$ is a finite quantity even in the
infinite limit, and the
computer calculations show small well structures in the rising branch of
the isospectral potentials and corresponding peaks in the wavefunctions.
The Bessel box is single-walled at the $\alpha$ limit.
For $I_0-K_0$, the integral is divergent in the infinite limit.
In this case the Bessel strictly isospectral box is like having walls of
different heights at the $c$ limit and $\alpha$ limit.
This peculiar box is a quite interesting object being the physical
embodiment of the strictly isospectral family of FRW potentials as a family
of Bessel solitonic potentials \cite{Keu} possessing translationally
moving structures. It can also be considered as a kind of cavity with
slowly moving walls and wells if $\lambda$ is varied even if the interval
$(-c, \alpha)$ is kept fixed.


In conclusion, we showed that the Hamiltonian constraint of canonical
quantum gravity for FRW minisuperspace
models may contain a one-parameter family of FRW cosmological potentials
and ``wavefunctions of the universe".
They are strictly isospectral (i.e. at zero energy)
WDW solutions of the FRW case including the
common solution in the limit $\lambda \rightarrow \pm\infty$.
Unfortunately, the strictly isospectral method can be applied only to either
pure one-dimensional cases as the FRW model here (see also \cite{rs})
or to separable cases \cite{taub}.


{\bf Acknowledgments}

This work was supported in part by the CONACyT Projects 4868-E9406
and 3898P-E9608.

\newpage
\centerline{
\epsfxsize=280pt
\epsfbox{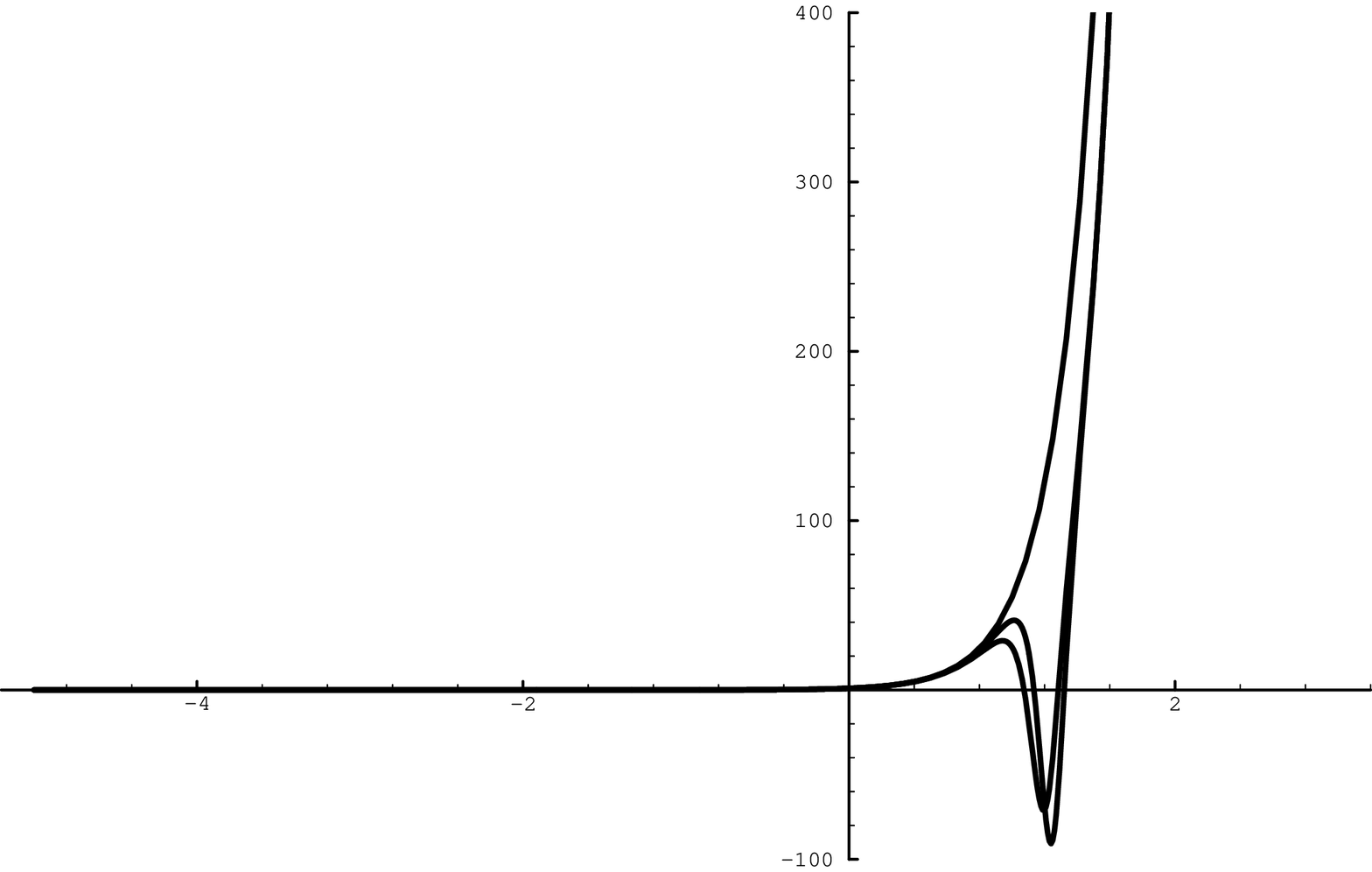}}
\vskip 4ex
\begin{center}
{\small{Fig. 1a}\\
The original potential and two members of the one-parameter strictly
isospectral family of FRW potentials in the case $Z_0=I_0$ for
$\lambda$ = 25 (small well) and 100 (bigger well)
}
\end{center}

\vskip 2ex
\centerline{
\epsfxsize=280pt
\epsfbox{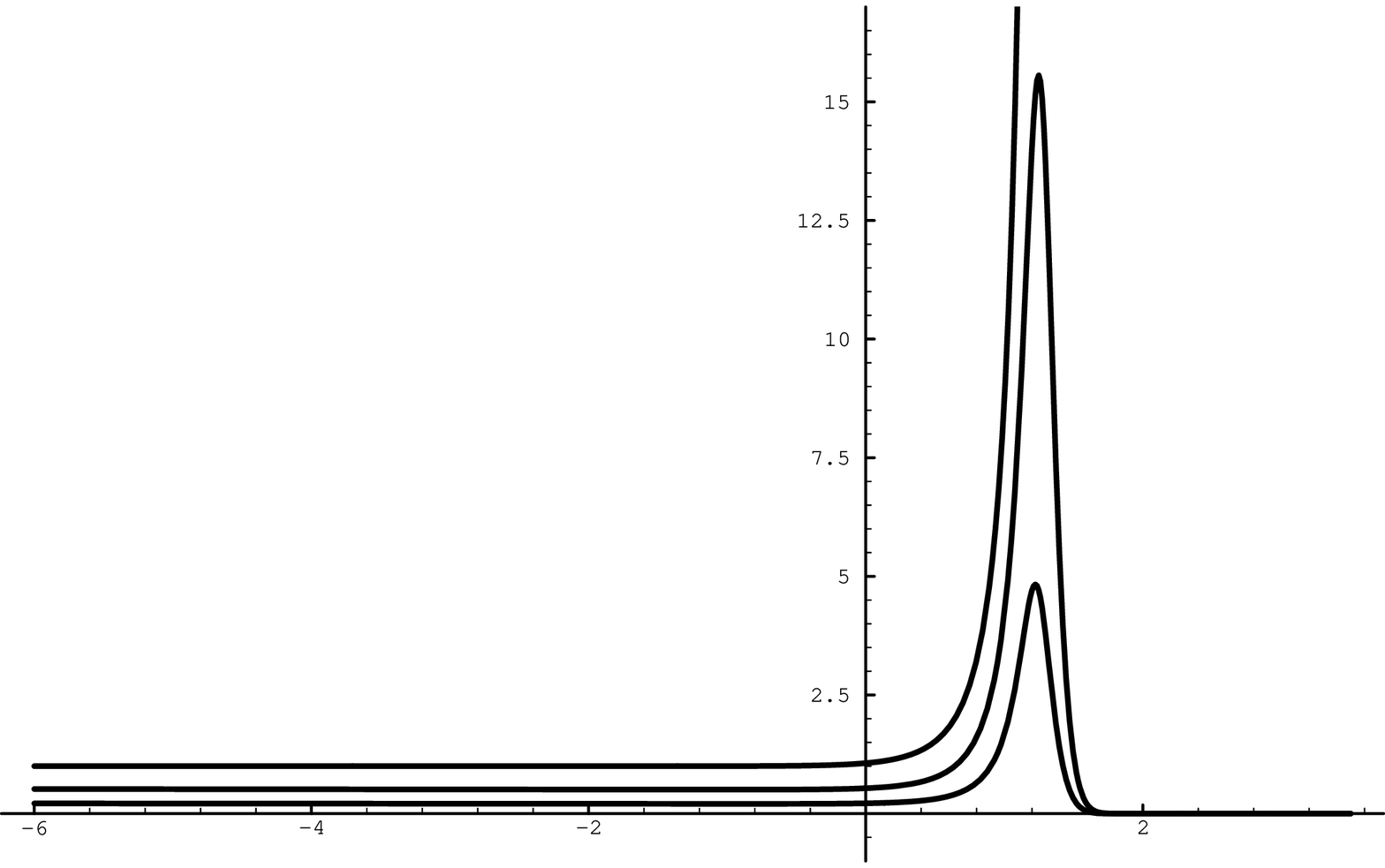}}
\vskip 4ex
\begin{center}
{\small{Fig. 1b}\\
The corresponding wavefunctions.
}
\end{center}

\vskip 2ex
\centerline{
\epsfxsize=280pt
\epsfbox{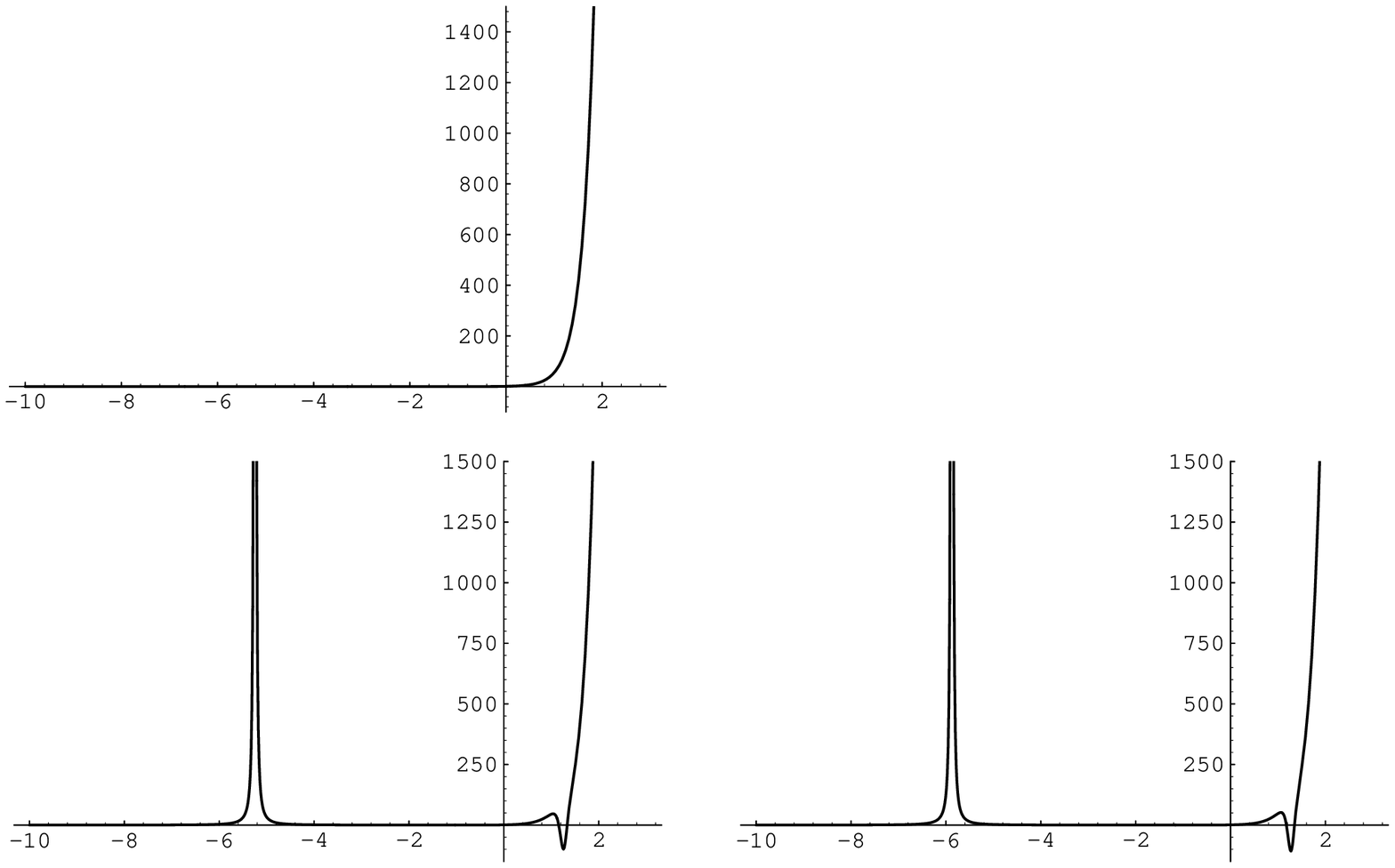}}
\vskip 4ex
\begin{center}
{\small{Fig. 2a}\\
The original potential and two members of the one-parameter strictly
isospectral family of FRW potentials for $I_{0}-K_{0}$,
and the same values of the
$\lambda$-parameter. On the arbitrary vertical scale the two
peaks close at a value of about 2,000.
}
\end{center}

\vskip 2ex
\centerline{
\epsfxsize=280pt
\epsfbox{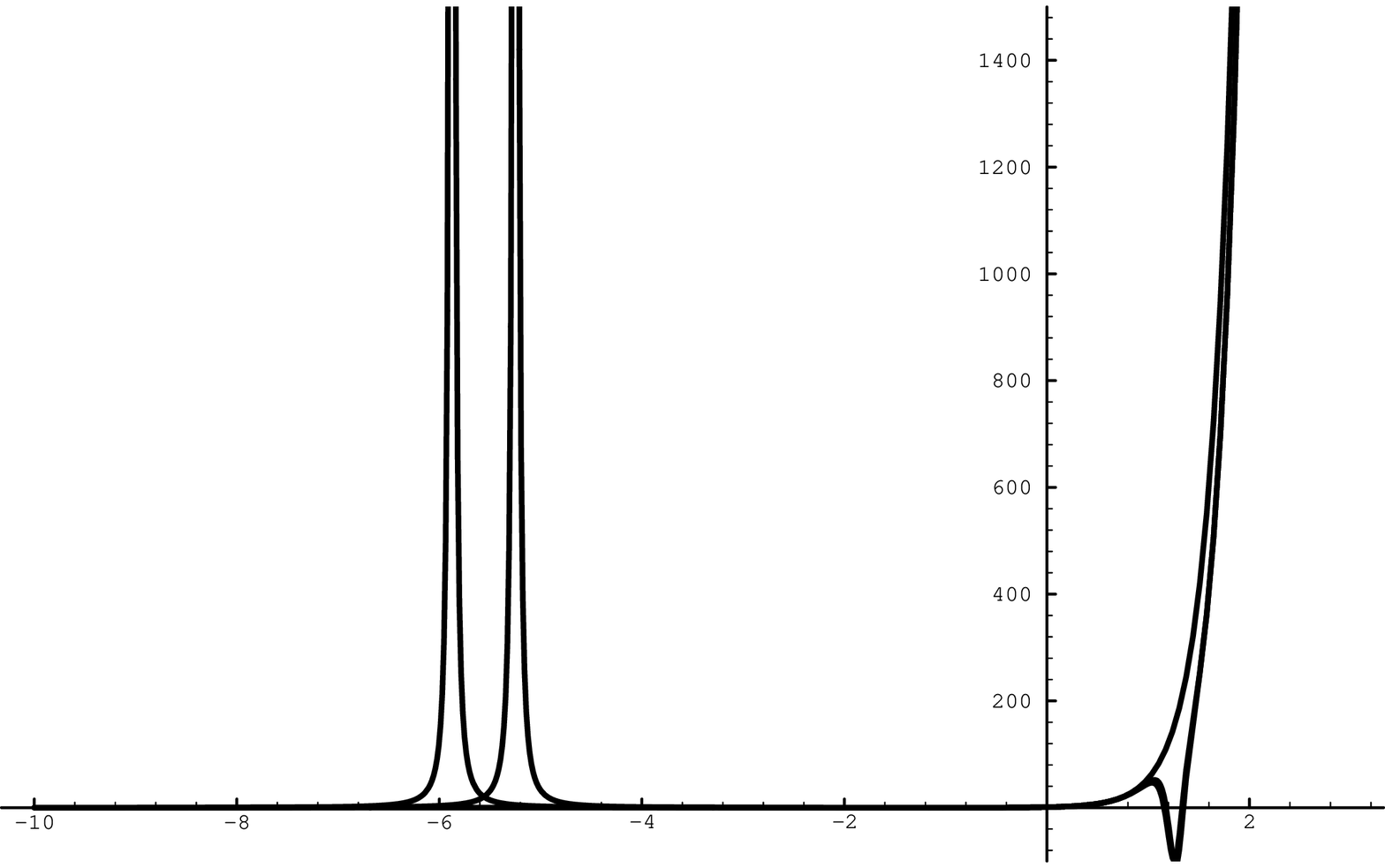}}
\vskip 4ex
\begin{center}
{\small{Fig. 2b}\\
All three potentials on the same plot, showing at the left the peak
corresponding to $\lambda =100$ and slightly at the right that for
$\lambda=25$. We have found that the peaks are moving to the left if
$\lambda$ is positive and increasing (i.e., beyond Pursey limit), while
they are moving toward the origin if $\lambda$ is less than the
Abraham-Moses limit and decreasing.
}
\end{center}

\vskip 2ex
\centerline{
\epsfxsize=280pt
\epsfbox{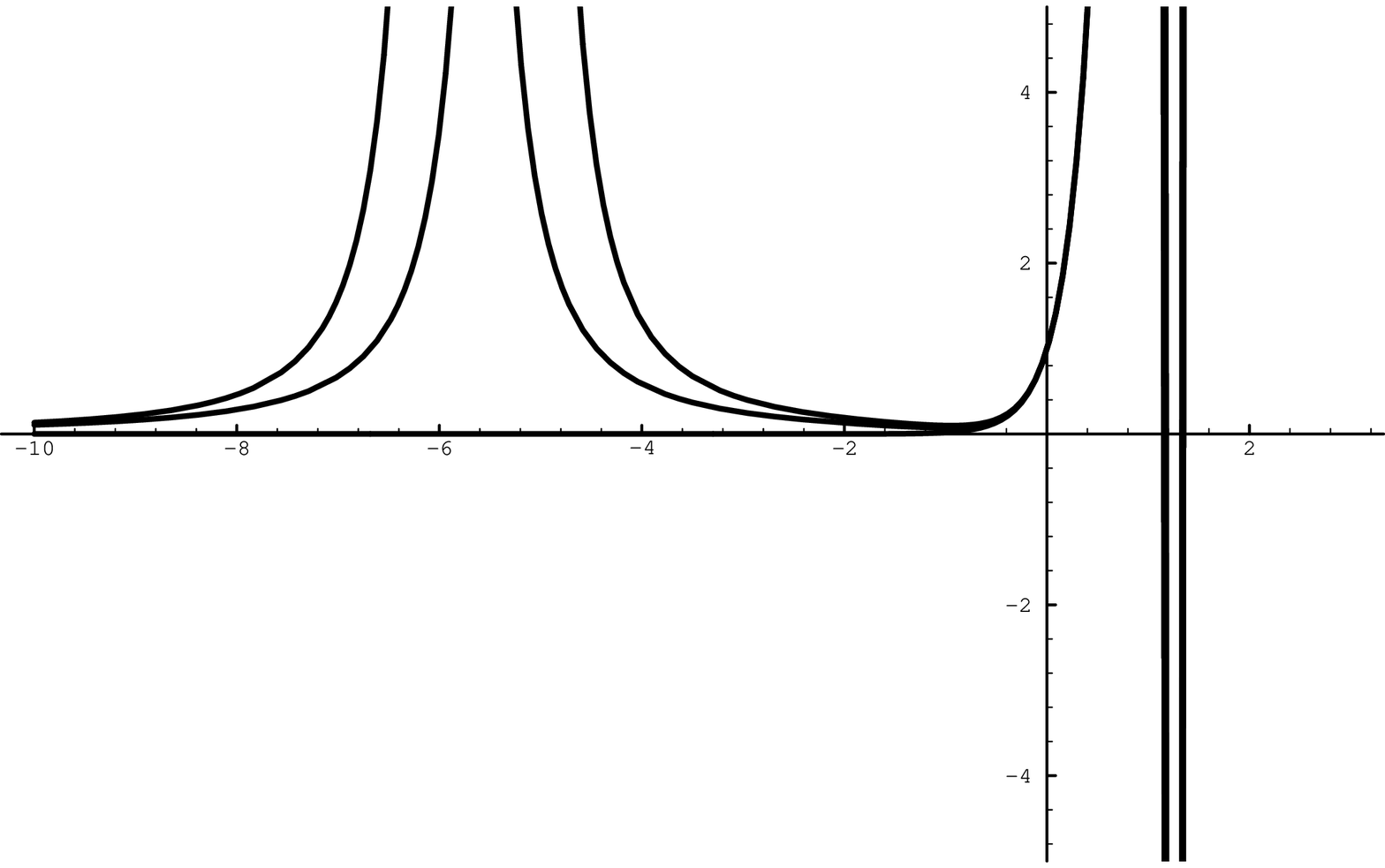}}
\vskip 4ex
\begin{center}
{\small{Fig. 2c}\\
A more detailed plot of the region close to the origin for the
three potentials showing the real trend of the plateau between the
small well structures at the right and the peaks at the left.
}
\end{center}

\vskip 2ex
\centerline{
\epsfxsize=280pt
\epsfbox{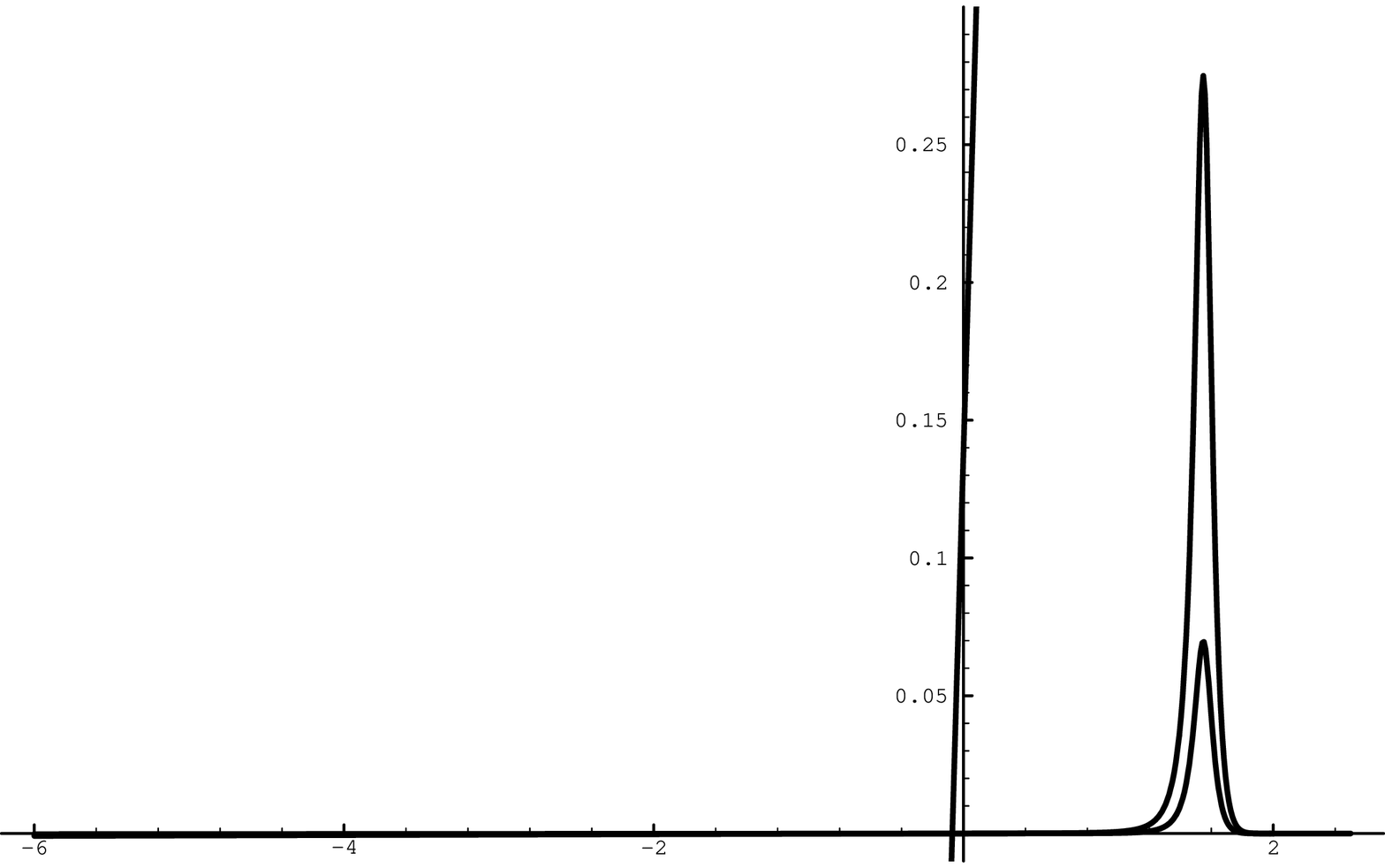}}
\vskip 4ex
\begin{center}
{\small{Fig. 3}\\
The strictly isospectral
`wavefunction of the universe', $Z_{gen}$, for $Z_0=I_0-K_0$ and the
$\lambda$ parameters
as in Figs. 1 and 2 (the small peak corresponds to $\lambda =25$, whereas
the big one is for $\lambda =100$). We notice also that the positive tails
(at the right of the peaks) are practically zero, while the negative
tails have a small positive slope. That implies that the wavefunctions are
not normalizable. The segment very close to the vertical axis is a portion
of the original `wavefunction', close to the value $I_{0}(1/2)-K_{0}(1/2)$.
}
\end{center}

\end{document}